\documentclass[twocolumn,aps,prl]{revtex4}
\usepackage{amsmath}
\usepackage{graphicx}
\usepackage{subfigure}
\usepackage[british]{babel}

\begin{document}

\title{On the dependency of rubber friction on the normal force or load: theory and experiment}
\author{G. Fortunato}
\affiliation{Bridgestone Technical Center Europe, Via del Fosso del Salceto 13/15 00128 Roma}
\author{V. Ciaravola}
\affiliation{Bridgestone Technical Center Europe, Via del Fosso del Salceto 13/15 00128 Roma}
\author{A. Furno}
\affiliation{Bridgestone Technical Center Europe, Via del Fosso del Salceto 13/15 00128 Roma}
\author{M. Scaraggi}
\affiliation{DII, Università del Salento, 73100 Monteroni-Lecce, Italy, EU}
\affiliation{PGI, FZ-J\"ulich, 52425 J\"ulich, Germany, EU}
\author{B. Lorenz}
\affiliation{PGI, FZ-J\"ulich, 52425 J\"ulich, Germany, EU}
\affiliation{www.MultiscaleConsulting.com}
\author{B.N.J. Persson}
\affiliation{PGI, FZ-J\"ulich, 52425 J\"ulich, Germany, EU}
\affiliation{www.MultiscaleConsulting.com}

\begin{abstract}
In rubber friction studies it is often observed that the kinetic friction coefficient $\mu$ depends on the nominal
contact pressure $p$. We discuss several possible origins of the pressure dependency of $\mu$: 
(a) saturation of the contact area (and friction force) due to high nominal squeezing pressure, 
(b) non-linear viscoelasticity,
(c) non-randomness in the surface topography, in particular the influence of the skewness of the surface roughness profile,
(d) adhesion, and (e) frictional heating. We show that
in most cases the non-linearity in the $\mu(p)$ relation is mainly due to process (e) (frictional heating), 
which softens the rubber, increases the area of contact, and (in most cases) reduces the viscoelastic contribution to the friction.
In fact, since the temperature distribution in the rubber at time $t$ depends on
on the sliding history (i.e., on the earlier time $t' < t$), 
the friction coefficient at time $t$ will also depend on the sliding history, i.e. 
it is, strictly speaking, a time integral operator.

The energy dissipation in the contact regions between solids in sliding contact
can result in high local temperatures which may strongly affect the area of real contact and the friction force (and the wear-rate). This is the case
for rubber sliding on road surfaces at speeds above $1 \ {\rm mm/s}$. In Ref. \cite{Fort} we have derived equations which
describe the frictional heating for solids with arbitrary thermal properties. 
In this paper the theory is applied to rubber friction on road surfaces. 
Numerical results are presented and compared to experimental data.
We observe good agreement between the calculated and measured temperature increase.
\end{abstract}

\maketitle

{\bf 1 Introduction}

The Coulomb friction law states that the friction force is proportional to the normal force or load, and is found to hold
remarkable well in many practical applications\cite{Rab,book}. The assumption that the friction coefficient is independent of the nominal contact pressure 
is often used also in rubber friction applications, e.g., 
in classical Finite Element Method (FEM) calculations, where the relation
is usually written as a relation between the local frictional shear stress $\tau (x,y)$ and the local contact pressure $p(x,y)$: 
$$\tau (x,y) = \mu p(x,y).\eqno(1)$$
Here it is implicitly assumed that the surfaces can be treated as smooth, but may have a macroscopic curvature (say with the radius of curvature $R$). 
This approach assumes that the longest wavelength roughness components on the surfaces
is short compared to the size of the nominal contact region, and also compared to the length scale $R$ associated with the macroscopic curvature. 
In that case the frictional shear stress $\tau(x,y)$ and contact pressure $p(x,y)$ in (1) must be considered as the locally average of the corresponding
microscopic quantities, which may vary rapidly in space down to length scales of order an atomic distance. 

In tire applications it is usually found that increasing the load results in a reduction of the effective friction 
coefficient (e.g., a reduction in the maximum of the $\mu$-slip curve). This is usually interpreted as a dependency of the friction coefficient
on the nominal contact pressure. However, increasing the normal load also changes the tire-road footprint, which affect the tire-road
friction. Thus, as will be shown in Sec. 7, an increase in the length of the tire-road footprint (as a result of an increase in the tire load),
usually results in a reduction of the maximum of the $\mu$-slip curve\cite{Tire1,Tire2}. 

In this paper we will discuss different processes which result in a pressure dependency of the friction coefficient $\mu$:
(a) saturation of the contact area (and friction force) due to high nominal squeezing pressure, 
(b) non-linear viscoelasticity,
(c) non-randomness in the surface topography, in particular the influence of skewness of the surface roughness profile,
(d) adhesion, and (e) frictional heating. We find that the latter mechanism is most important in tire applications and
in what follows we will focus mainly on this case.

When a rectangular block with nominally smooth surface is squeezed in contact with a nominally flat
substrate, because of surface roughness the area of real contact is usually only a very small fraction of the nominal contact area. For hard solids
the contact pressure in the area of real contact will therefore be very high. During sliding frictional energy dissipation will take place
in the area of real contact and because of the small volumes involved, at high enough sliding speed where thermal diffusion becomes
unimportant, the local (flash) temperatures may be very high. As a result local melting of the material, or other phase transformations, can take place.
In addition tribochemical reactions, and emission of photons or other particles, may occur at or in the vicinity of the contact regions. 
All these processes will also affect the friction force, e.g., if frictional melting occurs the melted film may act as a lubricant
and lower the friction as is the case, e.g., when sliding on ice or snow at high enough velocity. It is clear that a deep understanding of
the role of frictional heating is of crucial importance in many cases for understanding friction and wear processes.

Pioneering theoretical works on the temperature distribution in sliding contacts have been presented by Jaeger\cite{jaeg}, Archard\cite{Arch} 
and others\cite{other1,other2,other3,carb,Barber}.
In these studies a moving heat source is located at the sliding interface. However, some materials like rubber have internal friction and when such
solids are sliding on a rough surface frictional energy will be dissipated not just at the sliding interface but also some distance into the
viscoelastic material. The flash temperature effect related to this process was studied in Ref. \cite{PP1}, but neglecting the contribution from
the frictional interaction between the surfaces in the area of real contact, and also neglecting heat transfer to the substrate. 
In Ref. \cite{Trib} the theory of \cite{PP1} was extended to include these effects, but assuming that the substrate has infinite thermal conductivity.
In Ref. \cite{Fort} we remove this last restriction and present a general theory of frictional heating, which we will use in this paper (see Sec. 6).

There are many experimental studies of the temperature in frictional contacts, e.g., see Ref. \cite{Greg}. 
When analyzing experimental data it is usually assumed that the temperature is continuous at the rubber-substrate interface. However, the latter
assumption is in general not valid, in particular if surface roughness exist and the contact area is incomplete within the nominal 
contact region. One needs to use a heat transfer description\cite{heat,heat1,book,heatcom,Barber1,Green,mus,past} which relates the 
temperature jump $T_{\rm R}-T_{\rm S}$ between the rubber surface ($T_{\rm R}$)
and the substrate surface ($T_{\rm S}$) to the heat current $J$ through the interface via $J=\alpha (T_{\rm R}-T_{\rm S})$, 
where the heat transfer coefficient $\alpha$ in general depends on the sliding speed\cite{Trib}.
We used this more general approach in Ref. \cite{Fort} and also in this paper. 

\begin{figure}[htbp]
\centering
\includegraphics[width=0.35\textwidth]{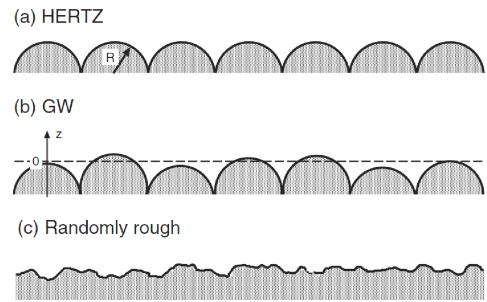}
\caption{Three different models of a ``rough'' surface. In case (a) all the asperities are equally high and have identical 
radius of curvature. Introducing asperities with a random height distribution as in (b) gives the Greenwood-Williamson approach 
towards contact mechanics. In (c) a real, randomly rough surface is shown, where the asperities are of different heights and curvature radii.} 
\label{hertz-gw-randomly_rough}
\end{figure}

\vskip 0.3cm
{\bf 2 Contact mechanics theory: short review}

Nearly all surfaces have roughness, usually extending from the linear size of the object down to atomistic length scales, say from cm to nm. 
This may involve $\sim 7$ decades in length scales corresponding to
$\sim 10^{21}$ degrees of freedom. The contact between solids with surface roughness cannot therefore in general be studied numerically on all relevant length scales 
without further simplifications. Thus, it is very important to develop analytical approaches to the contact between solids with rough surfaces.

Contact mechanics has a very long history\cite{[5]}. The first study
was presented by Hertz in 1882 ({\it \"Uber die Ber\"uhrung fester
elastischer K\"orper}), where he studied the frictionless
contact between elastic solids with smooth surface profiles
which could be approximated as parabolic close to
the contact area. This theory predicts a non-linear increase
of the contact area $A$ with the squeezing force $F$.
The most simple model of a rough surface consists of
a regular array of spherical bumps with equal radius of
curvature $R$ and equal height (see Fig. 1(a)). If such a
surface gets squeezed against an elastic solid with a flat
surface, one can approximately apply the Hertz contact
theory to each asperity. Thus, from this simple approach
one expects that the real area of contact will scale nonlinearly
with $F$. However, this is not in accordance with
experiments which show that the real area of contact is
proportional to $F$ as long as the contact area $A << A_0$,
where $A_0$ is the nominal contact area. This is also the
reason for why the friction force is usually proportional
to the load (Coulomb friction law).

Greenwood and Williamson (GW) proposed that the contact problem between two elastic
rough surfaces could be reduced to the problem of one
infinitely-hard rough surface acting on a flat elastic countersurface.
Within their model, the rough topography
was described by a large collection of hemispherical asperities
of uniform radius (which individually satisfied
the Hertzian approximation) with a height distribution
that followed a Gaussian law, see Fig. 1(b). This model
relies on the definition of ``asperity''. The asperity concept
itself has proven to be quite controversial and depends
on the resolution of the instrument used to measure
the surface profile. In addition, the long range elastic
coupling between the asperity contact regions is now known to strongly
influence contact mechanics. If an asperity is pushed
downwards at a certain location, the elastic deformation
field extends a long distance away from the asperity influencing
the contact involving other asperities further
away. This effect is neglected in the GW theory, significantly
limiting its prediction capabilities when applied
to most real surfaces. Additionally, in the GW model
the asperity contact regions are assumed to be circular
(or elliptical) while the actual contact regions (at high
enough experimental resolution) show fractal-like boundary
lines. Therefore, because of their complex geometries,
one should try to avoid explicitly invoking the nature
of the contact regions when searching for an analytical
methodology to solve the contact problem of two elastic
rough surfaces.

Recently, an analytical contact mechanics model that
does not use the asperity concept and becomes exact in
the limit of complete contact has been developed by Persson\cite{[6]}.
The theory accounts for surface roughness on all relevant
length scales and includes (in an approximate way)
the long range elastic coupling between asperity contact
regions. In this theory the information about the surface
enters via the surface roughness power spectrum, which
depends on all the surface roughness wavevectors components.

The contact mechanics formalism developed by Persson
\cite{[6]} is based on studying the interface between two
contacting solids at different magnifications $\zeta$. When the
system is studied at the magnification $\zeta$ it appears as if
the contact area equals $A(\zeta)$, but when the magnification
increases, it is observed that the contact is incomplete,
and the surfaces in the apparent contact area $A(\zeta)$ are
in fact only in partial contact, see Fig. 2. The theory
can be used to calculate the interfacial stress distribution
$P(\sigma,\zeta)$, from which one can obtain the area of real
contact as a function of the squeezing pressure $p$ and the
magnification $\zeta$. 
Furthermore, the theory predicts 
the average interfacial surface separation $\bar u$ (see Ref. \cite{[7]}),
and the
distribution of interfacial 
separation $P(u)$ (see Ref. \cite{[8]} and \cite{[9]}), which is very
important for the leak-rate and friction of seals\cite{add}.

\begin{figure}
\includegraphics[width=0.45\textwidth,angle=0]{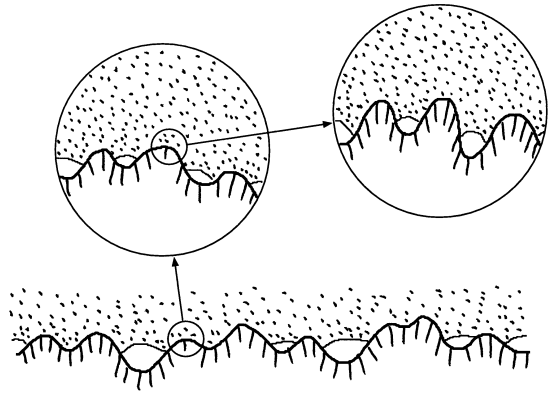}
\caption{
An elastic block (dotted area) in adhesive contact with a rigid rough substrate (dashed area).
The substrate has roughness on many different length scales, and the block makes partial contact
with the substrate on all length scales. When a contact area is studied, at low magnification
it appears as if complete contact occurs, but when the magnification is increased it is observed
that in reality only partial contact exists.
}
\label{1x}
\end{figure}

\begin{figure}
\includegraphics[width=0.9\columnwidth]{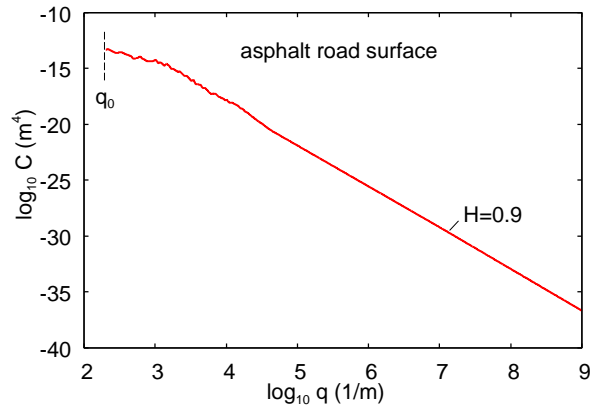}
\caption{\label{1logq.2logC.ps}
The surface roughness power spectrum used in the adhesion and contact mechanics calculations.
}
\end{figure}

\begin{figure}
\includegraphics[width=0.9\columnwidth]{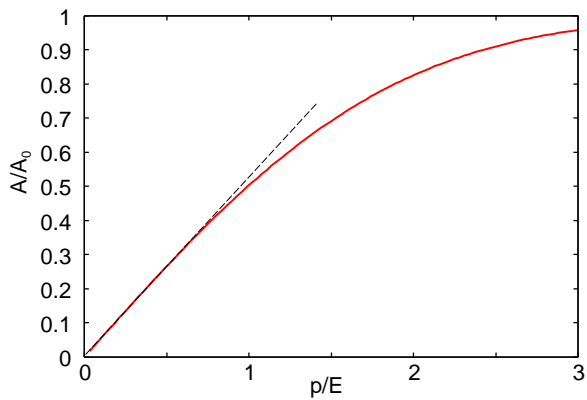}
\caption{\label{combineA.ps}
The area of real contact, as a function of the nominal contact pressure (in units of $E$), 
between an elastic solid with the Young's modulus $E$ and Poisson's ratio $\nu =0.5$,
and an asphalt road surface, which is considered rigid. 
}
\end{figure}

\vskip 0.3cm
{\bf 3 Friction between randomly rough surfaces}

Randomly rough surfaces can be generated mathematically by adding cosine
waves with different wavelength, suitable chosen amplitudes and with random phases.
Such surfaces have a Gaussian height distribution and, of course,
zero skewness. Real surfaces, e.g., produced by crack propagation or
sandblasting, are usually nearly randomly rough surfaces. 

Consider two elastic solids with randomly rough but nominal flat surfaces,
squeezed into contact with the nominal pressure $p$. In this case, assuming
linear elasticity, it has been shown both theoretically (using the Persson contact mechanics theory)
and in exact numerical studies that as long as the contact area $A$ is below $\sim 30\%$ of the nominal
contact area $A_0$, there is a linear relation between the relative contact
area $A/A_0$ and the squeezing pressure. 

The physical reason for this linear relation is as follows:
With increasing pressure $p$, existing contact
areas grow and new contact areas form in such a way that,
in the thermodynamic limit (infinitely large system), the (normalized)
interfacial stress distribution, and also the size distribution
of contact spots, are independent of the squeezing pressure.
From this it follows immediately that $A$
varies linearly with the squeezing force $pA_0$
The same linear scaling will be found for
any quantity that derives from the 
stress distributions, such as the elastic energy stored there.

The linear relation between $A$ and $p$ is consistent with the Coulomb friction
law which states that the friction force is proportional to the normal force or load:
If the area of real contact is proportional to the normal force, and the normalized size distribution of 
contact areas independent of the load, it follows immediately that the friction force is
proportional to the normal force.

Fig. \ref{combineA.ps}
shows the area of real contact between an elastic solid with the Young's modulus $E$ and Poisson's ratio $\nu =0.5$,
and an asphalt road surface. The solid line is 
obtained using the Persson contact mechanics theory.
For pressures up to $\approx E$ the area of real contact increases linearly with the pressure $p$, and we expect the same
for the rubber friction coefficient. For large pressures ($p > E$) the contact area approaches complete contact ($A/A_0=1$) and the
$A(p)$ curve flattens out or saturates.

The results presented above are for an elastic solids. However, we have recently shown
by exact numerical studies, and also using the Persson contact mechanics theory,
that very similar results prevail for viscoelastic solids (see also next section)\cite{Mich}.
For viscoelastic solids the viscoelastic modulus depends on the frequency, $E=E(\omega )$, and is a complex
quantity (where the imaginary part relates to energy dissipation). Since during sliding the
perturbing frequencies $\omega = qv$ (where $q$ is the wavenumber of a surface roughness component) increase with
increasing sliding speed $v$, the relation between
the contact area and the nominal contact pressure depends on the sliding speed (the contact area will decrease with increasing sliding speed). 
However, for any given sliding speed the area of real contact depends on the nominal contact pressure
in a very similar way as for elastic solids. 
Hence we conclude that for the nominal contact pressures of interest in tire applications
(typically $p < 0.6 \ {\rm MPa}$) a linear relation between contact area and the nominal contact pressure can be expected.

Rubber-like materials exhibit non-linear properties, and in particular filled
rubber compounds exhibit strain-softening associated with the break-up of the filler network. 
However, it is unlikely that this non-linearity
will result in a non-linear relation between $A$ and $p$ for small pressures. This follows
from the very general arguments given above, showing that
the (normalized) distribution of stresses at the interface does not change with increasing load.
Thus, most likely non-linearity in the viscoelastic properties will not 
change the linear relation between $p$ and $A$ found for linear viscoelastic solids.

\begin{figure}
\includegraphics[width=0.9\columnwidth]{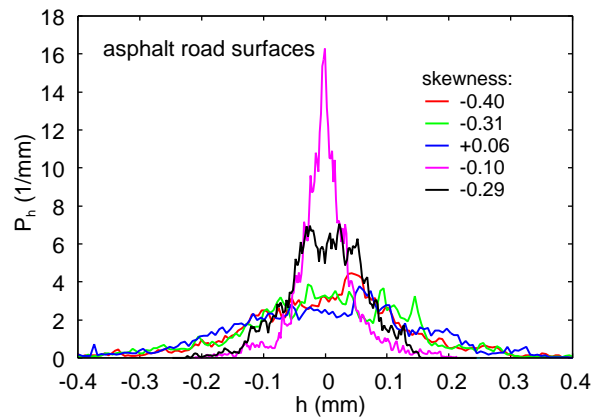}
\caption{\label{1h.2Ph.5types.ps}
The surface height probability distribution for 5 different asphalt road surfaces.
The average of the skewness over all 5 surfaces is $SK\approx -0.21$.
}
\end{figure}

\begin{figure}
\includegraphics[width=0.9\columnwidth]{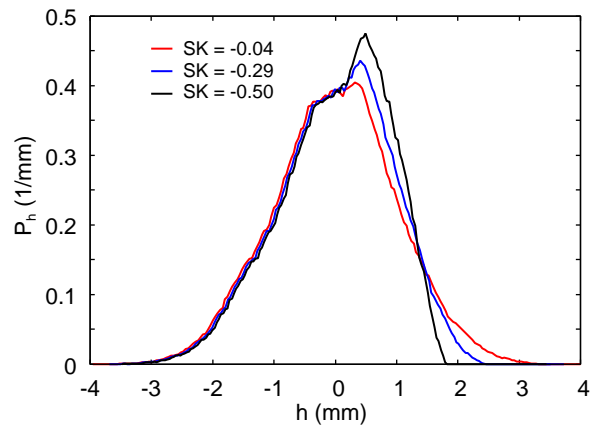}
\caption{\label{1h.2Ph.three.skewness.ps}
The surface height probability distribution for 3 mathematically generated (self-affine fractal) surfaces
with the same surface roughness power spectra.
}
\end{figure}

\begin{figure}
\subfigure[]{
\includegraphics[width=0.9\columnwidth]{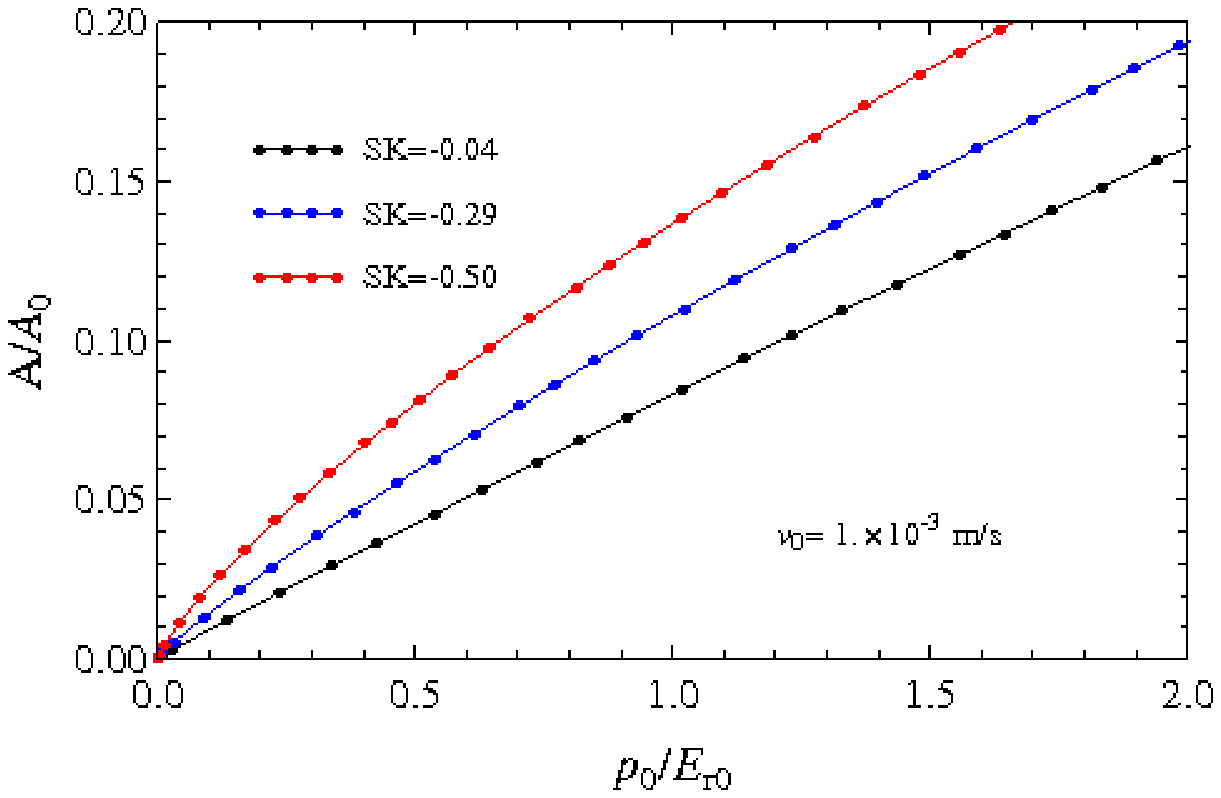} \label{1.all.sk.area.pressure.ps}
}
\\
\subfigure[]{\includegraphics[width=0.9\columnwidth]{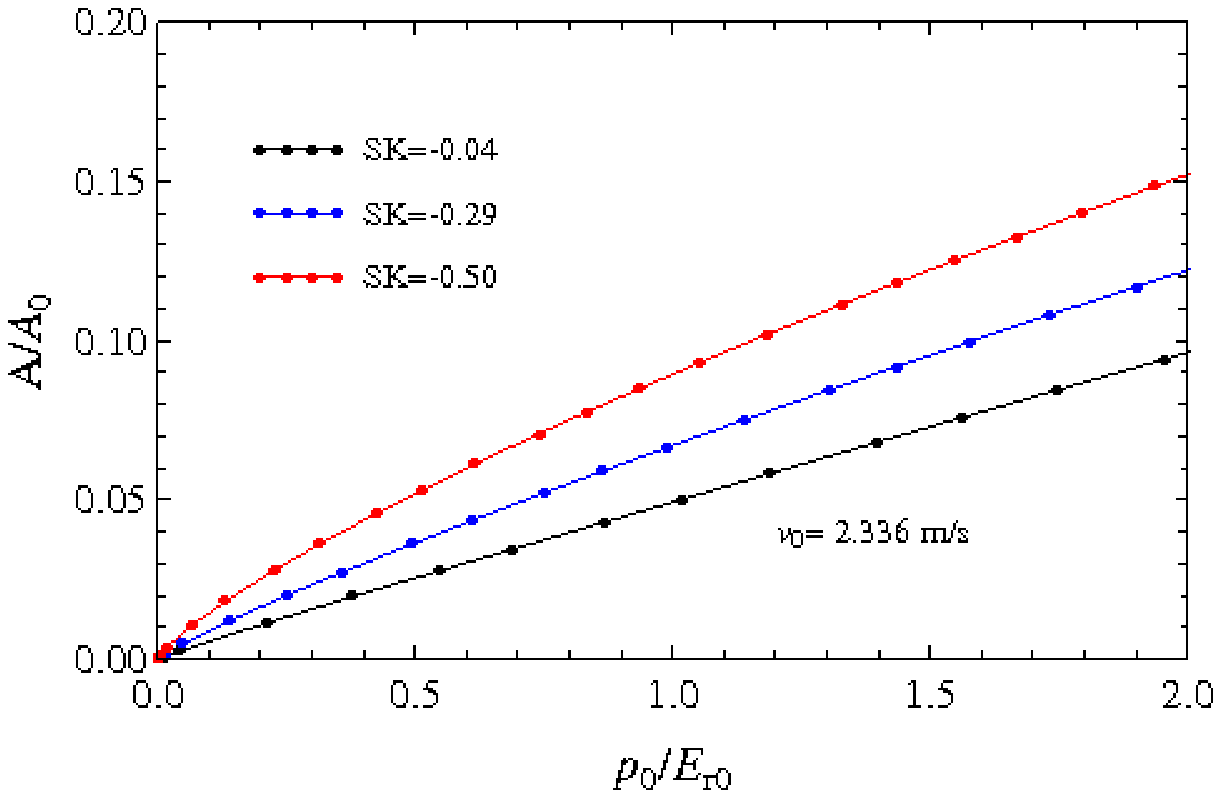} \label{3.all.sk.area.pressure.ps}
}
\caption{
The normalized contact area as a function of the squeezing pressure (in units of the
low frequency reduced modulus $E_{\rm{r0}}=E_0/(1-\nu^2)$) as obtained from numerical calculations.
Results are shown for three different surfaces with the same surface roughness
power spectrum but different skewness SK. The sliding speed a) $v=1 \ {\rm mm/s}$, b) $v=2.336 \ {\rm m/s}$.}
\label{numerics}
\end{figure}

\begin{figure}
\includegraphics[width=0.9\columnwidth]{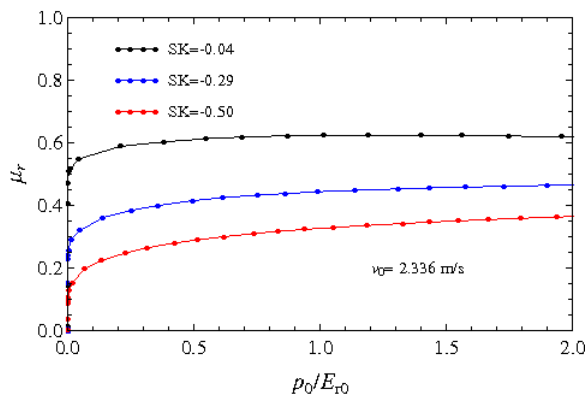}
\caption{\label{3.all.sk.rolling.pressure.ps}
The viscoelastic contribution to the friction coefficient as a function of the squeezing pressure (in units of the
low frequency reduced modulus $E_{\rm{r0}}=E_0/(1-\nu^2)$) as obtained from numerical calculations.
Results are shown for three different surfaces with the same surface roughness
power spectrum but different skewness SK. The sliding speed $v=2.336 \ {\rm m/s}$.}
\end{figure}

\vskip 0.3cm
{\bf 4 Friction on non-randomly rough surfaces}

The linear relation between contact area and the pressure expected for randomly rough surfaces
at small enough load (see above), will in general not hold for surfaces with non-random
roughness. This is clear already from a simple example: When a surface with a regular
arranged distribution of spherical bumps, all of equal radius and 
height [see Fig. \ref{hertz-gw-randomly_rough}(a)], is squeezed against a 
flat surface, according to the Hertz contact theory,
the contact area will depend non-linearly with the load as $A \sim p^{2/3}$. 

Asphalt road surfaces, at least when new (not extensively used), usually exhibit 
a negative skewness. This is a result of the production process where the 
hot asphalt is compressed and smoothed by rolling a heavy cylinder shaped
object on the asphalt surface. This is illustrated in Fig. \ref{1h.2Ph.5types.ps} which shows the
probability distribution of surface heights of 5 different asphalt road surfaces
as obtained from engineering stylus measurements. Most of the surfaces exhibit a negative skewness
with the average skewness $SK=-0.21$. Surfaces with skewness are non-random, and one cannot expect
such surfaces to exhibit a linear relation between $p$ and $A$. 

We have performed accurate numerical studies where we slide a rubber block with a flat
bottom surface against hard rough surfaces with negative skewness \cite{Mich,Michun}. We have prepared three 
different surfaces with the same (self-affine fractal) power spectra, but different
skewness as can be seen in Fig. \ref{1h.2Ph.three.skewness.ps}. 
The skewness was produced by scaling the roughness above the average surface plane with a scaling
factor $<1$ to get negative skewness. In the calculations
we use the viscoelastic modulus measured for a tread rubber compound, see Ref. \cite{Fort}.

Fig. \ref{1.all.sk.area.pressure.ps}
shows the normalized contact area as a function of the squeezing pressure (in units of the
low frequency modulus $E_{\rm{r0}}$) as obtained from the numerical calculations.
Results are shown for the three different surfaces with the same surface roughness
power spectrum but different skewness. The sliding speed $v=1 \ {\rm mm/s}$.

Fig. \ref{3.all.sk.area.pressure.ps} shows similar results as in Fig. \ref{1.all.sk.area.pressure.ps} but for
a higher sliding speed, $v \approx 2.3 \ {\rm m/s}$.
Note that the skewness results in some non-linearity in the $A(p)$ relation for small contact pressure, and also that the contact
area is larger with the skewness as compared to the case of (nearly) vanishing skewness (black data points).
Both effects are easy to understand: The negative skewness implies that the surface roughness above the
average plane is reduced in the height as compared to the surface with vanishing skewness. This will, for a given
contact pressure, result in a larger contact area. The non-linearity in the $A(p)$ relation for small $p$
results from the fact that at higher contact
pressures the rubber penetrates deeper into the surface profile and will experience the larger roughness which
exists deeper into the surface profile.

Note also that the contact area is smaller at the higher sliding speed. This is due to the higher perturbing
frequencies $\omega = qv$ (see Sec. 3) acting on the rubber surface from the road asperities
and to the fact that the magnitude of the viscoelastic modulus $|E(\omega )|$ increases with increasing $\omega$. 

Fig. \ref{3.all.sk.rolling.pressure.ps}
shows the viscoelastic contribution to the friction coefficient as a function of the squeezing pressure (in units of the
low frequency reduced modulus $E_{\rm{r0}}=E_0/(1-\nu^2)$) as obtained from the exact numerical calculations.
Results are shown for the same surfaces as in Fig. \ref{1.all.sk.area.pressure.ps} and for the sliding speed $v=2.336 \ {\rm m/s}$.
The decrease in the friction coefficient with decreasing nominal contact pressure for small contact pressures
is a finite-size effect, and for larger system the friction coefficient approach a constant value for small contact pressures (see Ref. \cite{Mich,Michun}).
Note that with increasing magnitude of the (negative) skewness the friction decreases. 
This is again due to the fact that the surfaces above the average plane becomes smoother
as the magnitude of the (negative) skewness becomes larger.

We finally note that for real asphalt road surfaces, the skewness of the surface may have much smaller
influence on the contact area (and the friction) than found in the model study above. The reason for this is as follows:
In our study we generate the roughness profile by scaling the roughness above the average plane on a randomly rough
surface (with skewness $SK=0$) with a factor $<1$. This will scale all asperities (big or small) with the same factor.
However, asphalt road surfaces are produced by mixing stone particles of different sizes with a binder. Since the
flattening of the hot asphalt, by rolling a cylinder body on top of it, will not break the stone particles
it is likely that the short wavelength roughness resulting from the small stone particles
will not be modified in the region above the average plane as compared to below the average plane.
Only the position of the big stone particles will be modified such as to result in a smoother surface after
the rolling action. But both the contact area and the rubber friction depends mainly on the shorter wavelength roughness,
and will not be much affected by modification of the long wavelength roughness due to the flattening.

\begin{figure}
\includegraphics[width=0.9\columnwidth]{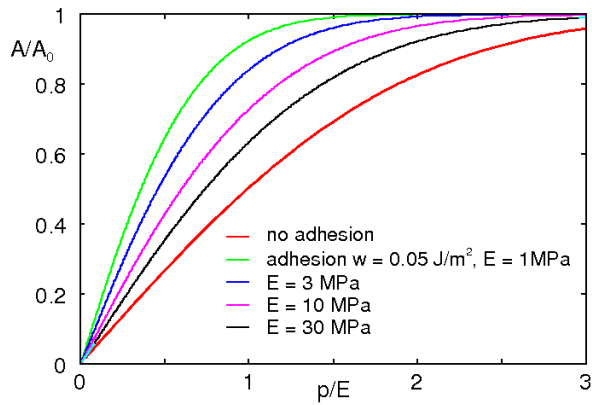}
\caption{\label{1p.over.E.2Area.ps}
The area of real contact between an elastic solid with the Young's modulus $E$ and Poisson's ratio $\nu =0.5$,
and a (rigid) asphalt road surface. The black line is without adhesion, and the other lines with adhesion using the work
of adhesion between flat surfaces $w=0.05 \ {\rm J/m^2}$, as is typical for the (adiabatic) work of adhesion
between rubber and many solids (in this case the physical origin of $w$ is usually due to the
weak Van der Waals interaction).
}
\end{figure}

\vskip 0.3cm
{\bf 5 Rubber friction: role of adhesion}

We now discuss the influence of adhesion on rubber friction and the contact area.
Fig. \ref{1p.over.E.2Area.ps}
shows the area of real contact between an elastic solid with the Young's modulus $E$ and Poisson's ratio $\nu =0.5$,
and an asphalt road surface with the surface roughness power spectrum $C(q)$ given in Fig. \ref{1logq.2logC.ps}. 
The black line is without adhesion [from Fig. \ref{combineA.ps}(a)], and the other lines with adhesion using the work
of adhesion between flat surfaces $w=0.05 \ {\rm J/m^2}$, as is typical for the (adiabatic) work of adhesion
between rubber and many solids (in this case the physical origin of $w$ is usually due to the
weak Van der Waals interaction). Note that even when adhesion is included the contact area increases linearly
with the nominal contact pressure, but with a slope which is larger than in the absence of adhesion. 

For smoother surfaces than used above the contact area for vanishing applied load (or nominal contact pressure) may be
non-zero, and the $A(p)$ relation non-linear already for small applied load. In this case
the friction coefficient will depend on the load and, in particular, increase towards infinite as the load decreases towards zero.
This has been observed in many experiments, but for
tread rubber in contact with clean road surfaces this will not be the case because of the large surface roughness. We conclude that
in tire applications, at least on clean surfaces,
including adhesion will still result in a friction coefficient which is independent of the nominal contact pressure for small enough
contact pressures. Nevertheless, the adhesive interaction will increase the area of real contact and the friction force.

\begin{figure}
\includegraphics[width=0.8\columnwidth]{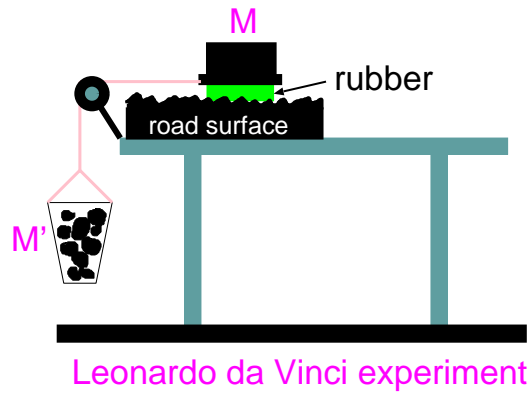}
\caption{\label{Leanardo.ps}
Simple friction tester (schematic) used for obtaining the friction coefficient $\mu = M'/M$ as a function of the sliding speed. The sliding distance is measured using a distance sensor and the sliding velocity obtained by dividing the sliding distance with the sliding time. This set-up can only measure the friction coefficient on the branch of the $\mu(v)$-curve where the friction coefficient increases with increasing sliding speed $v$.
}
\end{figure}

\begin{figure}
\includegraphics[width=0.9\columnwidth]{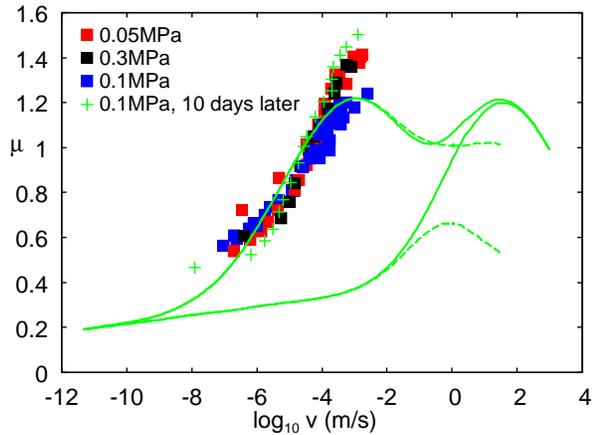}
\caption{\label{1logv.2mu.compoundC.concrete.ps}
The measured and calculated friction coefficient as a function of sliding speed for a rubber compound 
on concrete surface. The upper green lines are the total
calculated rubber friction coefficient and the lower green lines the viscoelastic contribution.
The solid lines are without flash temperature and the dashed lines with the flash temperature.
The background temperature is $T_0 = 8^\circ {\rm C}$
and the nominal contact pressure $p=0.05 \ {\rm MPa}$ (red squares), $p=0.3 \ {\rm MPa}$ (black squares)
and $p=0.1 \ {\rm MPa}$ (blue squares and green +).
}
\end{figure}

\begin{figure}
\includegraphics[width=0.9\columnwidth]{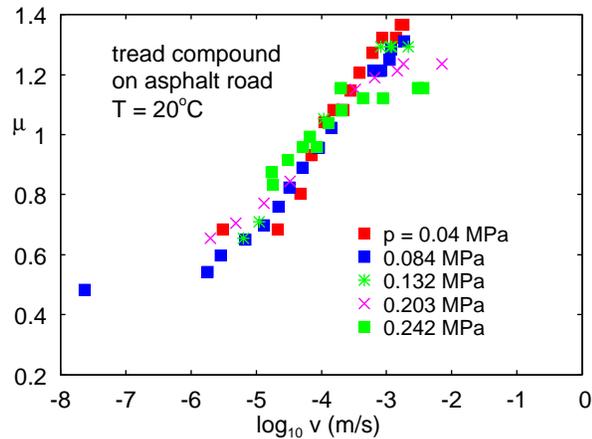}
\caption{\label{1logv.2mu.many.load.new.ps}
The measured friction coefficient as a function of sliding speed for a rubber compound 
on an asphalt road surface. 
The background temperature is $T_0 = 20^\circ {\rm C}$, and for the nominal contact pressure indicated
in the figure.
}
\end{figure}

\vskip 0.3cm
{\bf 6 Rubber friction: experimental results for low sliding speed}

At low sliding speed (negligible frictional heating), many experiments show that the rubber friction coefficient 
on rough surfaces is approximately independent of the normal load.
We now report on measured rubber friction for tread rubber compounds sliding against concrete and asphalt road surfaces. The measurements were performed using 
the Leonardo da Vince set-up shown in Fig. \ref{Leanardo.ps}. The slider consists of two rubber blocks glued to a wood plate. 
This simple friction tester can be used for obtaining the friction coefficient $\mu = M'/M$ as a function of the sliding speed. 
The sliding distance is measured using a distance sensor, and the sliding velocity obtained by dividing the sliding distance with the sliding time. \
This set-up can only measure the friction coefficient on the branch of the $\mu(v)$-curve where the friction coefficient increases with increasing sliding speed $v$,
and the maximum velocity data points, in the measured data presented below, correspond to local maximum in the $\mu (v)$ curves.

Fig. \ref{1logv.2mu.compoundC.concrete.ps}
shows the measured and calculated friction coefficient as a function of sliding speed for a rubber tread compound 
on concrete (see Ref. \cite{Hank} for details related to theory prediction). 
The nominal contact pressure $p=0.05 \ {\rm MPa}$ (red squares), $p=0.3 \ {\rm MPa}$ (black squares),
and $p=0.1 \ {\rm MPa}$ (blue squares and green +). The green + data was measured 10 days after the black-square measurement.
During this time period the road surface and the rubber slider were both kept in the normal atmosphere. Clearly a modification
of the rubber (and/or road) surface properties must have taken place. However, within the
accuracy of the measurements, there is no dependency of the friction coefficient on the normal load, when the experiments
are performed on surfaces the same day.

Fig. \ref{1logv.2mu.many.load.new.ps}
shows the measured friction coefficient as a function of sliding 
speed for a another rubber tread compound 
on an asphalt road surface. 
The results are for the background temperature is $T_0 = 20^\circ {\rm C}$,
and for several nominal contact pressure is indicated in the figure.
For this surface it appears as if the maximum of the friction coefficient decreases with
increasing nominal contact pressure.
At the same time the friction coefficient for lower velocities may be slightly increased
when the nominal contact pressure increases. 

We now discuss why the friction coefficient in Fig. \ref{1logv.2mu.compoundC.concrete.ps} may depend on the nominal contact pressure
for sliding speeds $v < v_{\rm max}$, close to the sliding speed $v=v_{\rm max}$ 
where the friction coefficient is maximal. We assume that the contact area is proportional to the 
nominal contact pressure $p$, and that with increasing pressure $p$, existing contact
areas grow and new contact areas form in such a way that the (normalized)
interfacial stress distribution, and also the size distribution
of contact spots, are independent of the squeezing pressure (see Sec. 3).
In this case the only thing which could influence the sliding friction is the concentration
of macroasperity contact regions which increase proportional to $p$.

We suggest the following picture: for velocities well below the velocity $v_{\rm max}$, where the friction coefficient is
maximal in Fig. \ref{1logv.2mu.many.load.new.ps}, the rubber in all the macroasperity contact regions slip relative to the substrate
with the same velocity $v$ as the upper surface of the rubber block. In that case the elastic (or rather viscoelastic) 
lateral coupling between the rubber contact regions, arising from the rubber deformation field around the asperity contact regions, is not changing 
in time, and it is basically irrelevant for the friction. 
Now, for velocities $v>v_{\rm max}$ the rubber friction decreases with increasing sliding speed. In the present case where the driving force is constant,
this results in an accelerated motion of the rubber block
when the driving force becomes larger than what is needed to reach the maximum in the friction coefficient. If instead the upper surface of the block
would be driven with a constant velocity $v>v_{\rm max}$ the bottom surface of the block would perform stick-slip motion. But with the same argument one
expect the rubber in the macroasperity contact regions to perform stick-slip motion for $v> v_{\rm max}$. However, due to the stochastic fluctuations in the nature of
the roughness in the macroasperity contact regions one expect not a sharp onset velocity for stick-slip motion at the macroasperity level, but a distribution
of onset velocities. Thus we expect that close to the friction maximum, but already for $v < v_{\rm max}$, the individual contact regions perform stick-slip
motion. In this case the shear deformation field of a macroasperity contact region depends on the shear deformation field of a nearby macroasperity contact regions.
Thus, we expect some correlation in the local stick-slip events when the velocity is close to the point where the friction coefficient is maximal.
For example, if the rubber in a macroasperity contact region slip into a state where the shear stress vanish, the tangential force lost in this contact region will
distribute itself on the nearby rubber macroasperity contact regions, where the shear stress now may increase to the point of resulting in
local slip, and so on.
Clearly, this lateral coupling, and the way the stress redistribute itself in response to a local slip at a macroasperity contact region,
will depend on the average separation between the macroasperity contact regions, and hence on the 
concentration of the macroasperity contact regions, which increases with increasing
nominal contact pressure. The stick-slip events should manifest itself in the power spectrum of the block velocity or perhaps
in the acoustic power spectrum, so studying these quantities should be one way to test the hypothesis presented above.

\begin{figure}
\includegraphics[width=1.0\columnwidth]{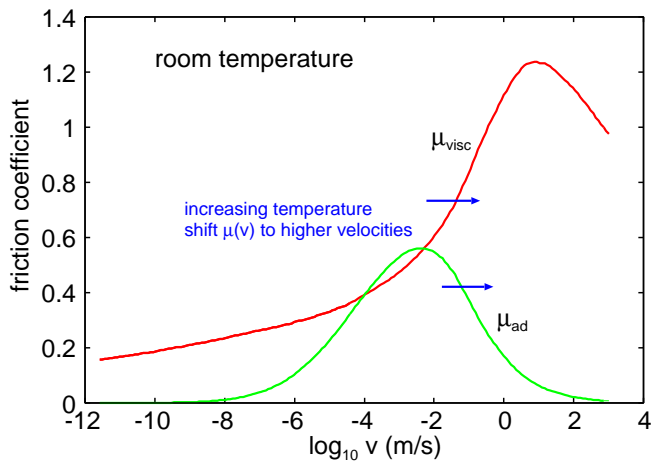}
\caption{\label{mu.ad.mu.visc.T.ps}
Schematic picture illustrating that an increase in the temperature shifts both $\mu_{\rm cont}(v)$ and $\mu_{\rm visc}(v)$ towards higher sliding speeds.
}
\end{figure}

\begin{figure}
\includegraphics[width=0.8\columnwidth]{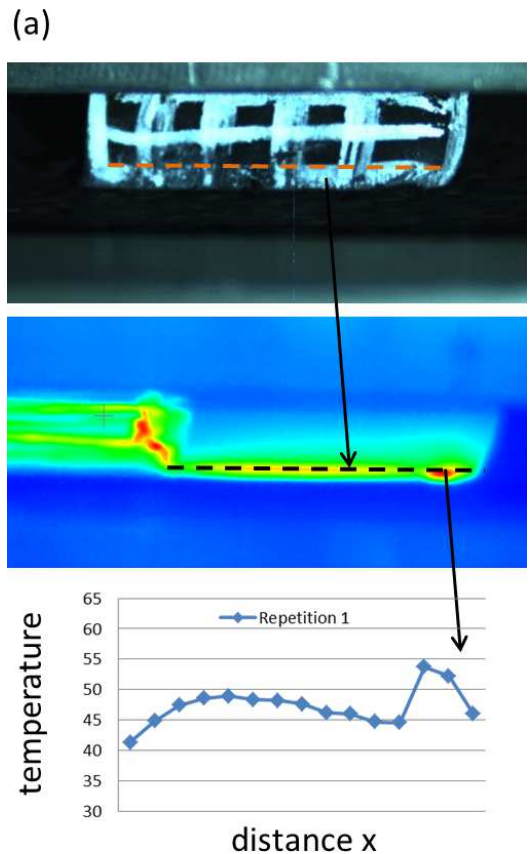}
\caption{\label{Slide_Temperature_profile_1.ps}
The measured temperature profile before run-in, after sliding $s=3.5 \ {\rm m}$. The rubber block
is $L=2.5 \ {\rm cm}$ long in the sliding direction and $0.7 \ {\rm cm}$ high.
}
\end{figure}

\begin{figure}
\includegraphics[width=0.8\columnwidth]{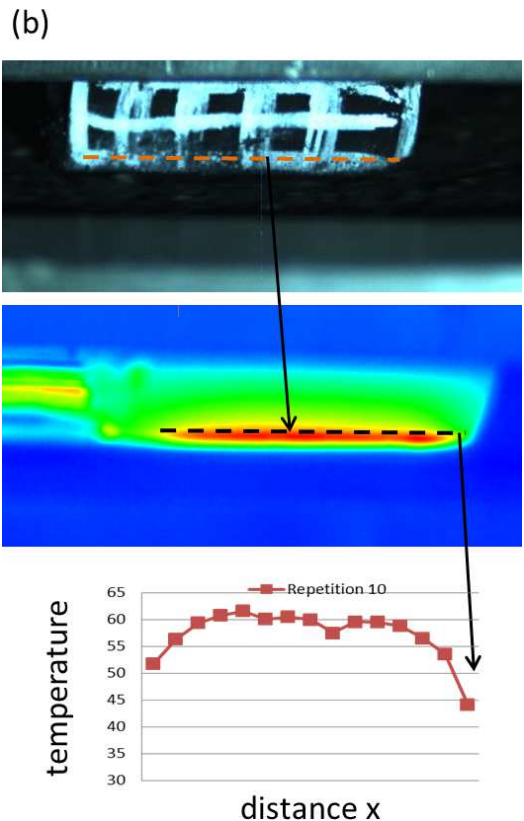}
\caption{\label{Slide_Temperature_profile_2.ps}
The measured temperature profile $T(x)$ after run-in. Run-in consists of 10 repetitions, each involving sliding of $s=3.5 \ {\rm m}$, i.e., a total distance of $35 \ {\rm m}$.
Between each run is 10 second waiting time. The temperature profile is after the last repetition.
}
\end{figure}

\begin{figure}
\includegraphics[width=0.9\columnwidth]{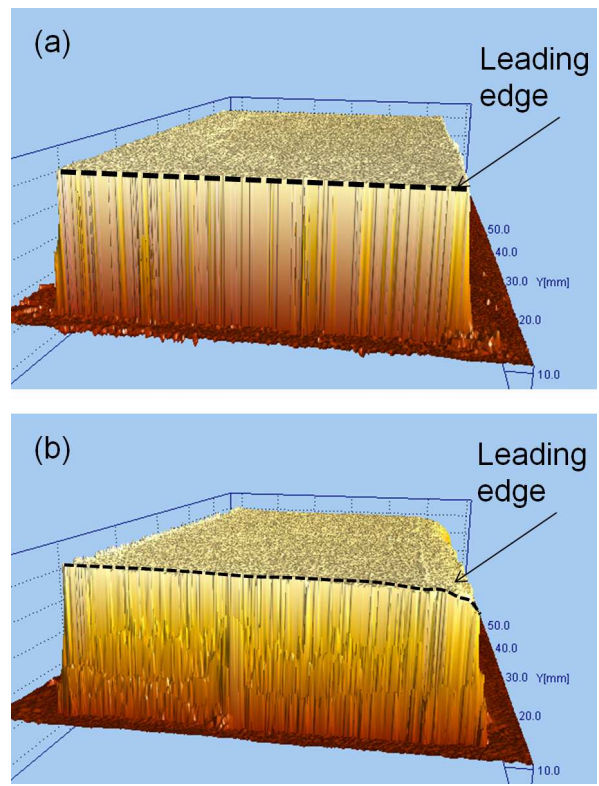}
\caption{\label{Slide_Laser_Profile_Rectangular-1a_and_2a.ps}
The measured temperature profile $T(x)$ of the rubber block (a) before run-in,
(b) after run-in (10 repetitions, each involving sliding $s=3.5 \ {\rm m}$).
}
\end{figure}

\begin{figure}
\includegraphics[width=0.9\columnwidth]{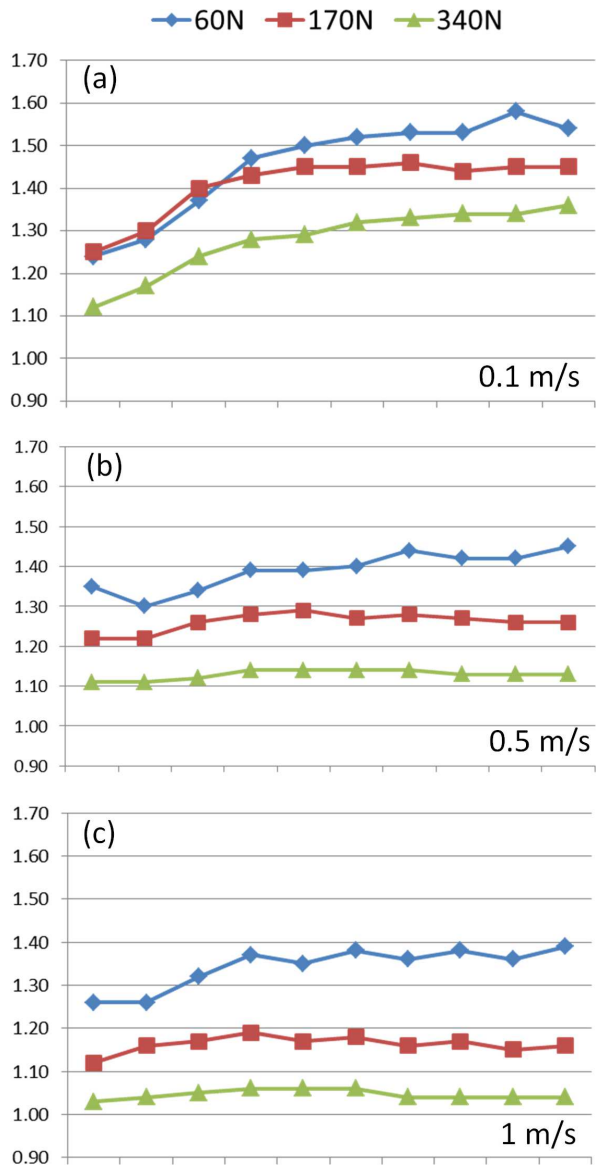}
\caption{\label{page4_all.ps}
The measured friction coefficients as a function of the number
of repetitions during run-in. Each repetition involves sliding
3.5 m at the sliding speeds (a) $v= 0.1$, (b) 0.5, and (c) $1 \ {\rm m/s}$.
Results are shown for the nominal contact pressures $0.05$ (blue lines),
$0.15$ (red), and $0.3 \ {\rm MPa}$ (green).
}
\end{figure}

\begin{figure}
\includegraphics[width=0.9\columnwidth]{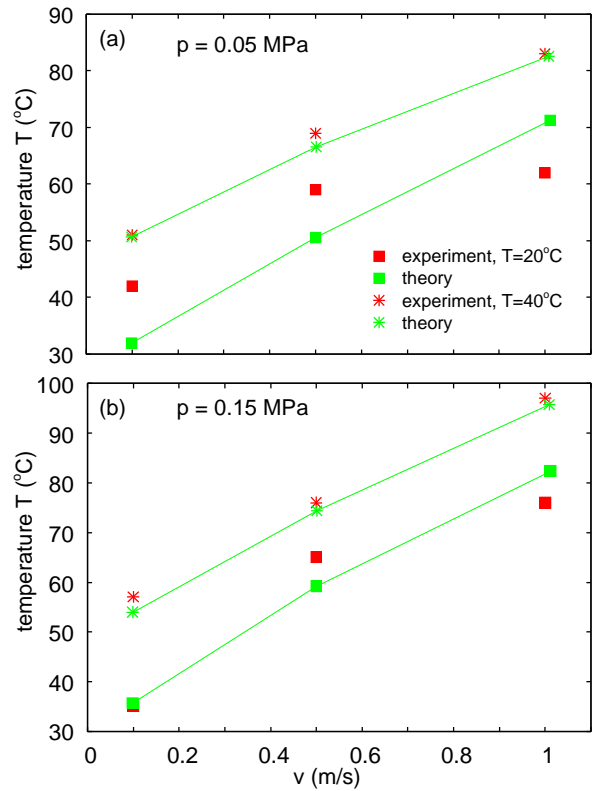}
\caption{\label{1v.2T.p=0.05MPa.p=0.15.ps}
The measured maximal temperature on the rubber side wall (red data points) and the calculated
average (over the $y$-direction) rubber temperature at $x=L/2$ at rubber-road interface (green symbols and lines).
Results are shown when the initial temperature equals $T=20$ and $40^\circ {\rm C}$, and for the contact
pressures (a) $p=0.05$ and (b) $0.15 \ {\rm MPa}$. The block is $L=2.5 \ {\rm cm}$ long in the sliding direction
and the temperature refers to a sliding distance $s=3.5 \ {\rm m}$.
}
\end{figure}

\begin{figure}
\includegraphics[width=0.9\columnwidth]{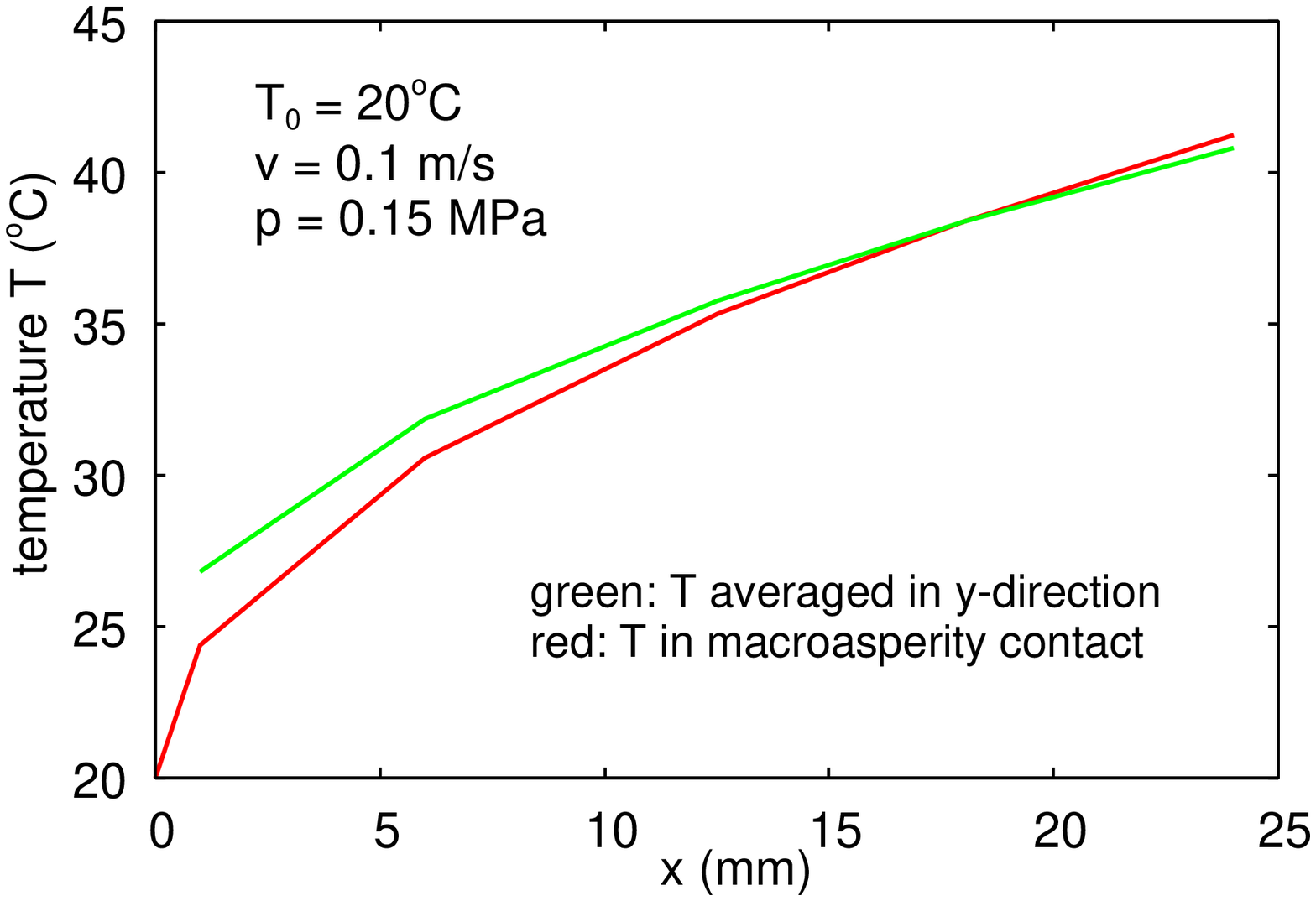}
\caption{\label{1x.2T.v=0.1m.per.s.ps}
The calculated temperature $T(x)$ as a function of the position $x$ along the sliding
direction with $x=0$ at the leading edge and $x=L=2.5 \ {\rm cm}$ at the trailing edge. 
The temperature profile is after sliding $s=3.5 \ {\rm m}$ at the nominal contact pressure
$p=0.15 \ {\rm MPa}$ and with the initial temperature $T=20^\circ {\rm C}$. The red line is the temperature
in the macroasperity contact regions and the green line the temperature averaged 
over the $y$-direction orthogonal to the sliding direction.
}
\end{figure}

\vskip 0.3cm
{\bf 7 Rubber friction: role of frictional heating}

There are two contributions to rubber friction, one from the viscoelastic deformations of the rubber by the road asperities\cite{[6],Klu}, and another
from shearing the area of real contact\cite{Shall,PersVol}, the latter is usually referred to as the adhesive contribution\cite{Hank}.
Earlier studies have shown that at room temperature the maximum in the adhesive contribution is located below the typical slip velocities in
tire applications ($1-10 \ {\rm m/s}$), while the maximum in the viscoelastic contribution may be located above typical sliding speeds as is indicated in Fig. \ref{mu.ad.mu.visc.T.ps}.
Increasing the temperature shifts both $\mu_{\rm cont}(v)$ and $\mu_{\rm visc}(v)$ towards higher sliding speeds, and also increases the area of real contact
$A$, making the adhesive contribution more important.
Depending on the relative importance of the adhesive and viscoelastic contribution to the friction, the friction coefficient
may increase or decrease with incresasing temperatures. For passenger car tires at typical operating temperatures it appears as if the friction
usually decreases with increasing temperature while for special tires, e.g., motorsport tires, the friction may increase as the temperature increases up to
rather high temperatures.

The energy dissipation in the contact regions between solids in sliding contact
can result in high local temperatures which may strongly affect the contact area and the friction. This is the case
for rubber sliding on road surfaces at speeds above $1 \ {\rm mm/s}$. In Ref. \cite{Fort} we have derived equations which
describe the frictional heating for solids with arbitrary thermal properties. 
Here we apply this theory to rubber friction on road surfaces, and we take into account
that the frictional energy is partly produced inside the rubber due to the internal friction
of rubber, and partly in a thin (nanometer) interfacial layer at the 
rubber-road contact region. The heat transfer between the rubber and
the road surface is described by a heat transfer coefficient which depends on the
sliding speed. In most cases this heat transfer coefficient is so large that the temperature basically is
continuous in the contact region at the rubber-road interface, and we have made this assumption here too.

At the Bridgestone lab we have developed a rubber friction tester where a rubber block
is slid on a circular asphalt road track. Using this new set-up we have studied the temperature distribution
on the road surface behind the rubber block using an infrared camera.

We have found that a relatively long run-in phase is necessary in order to obtain reproducible data. During run-in
the shape of the rubber block changes due to wear which results in a more uniform (nominal) contact pressure distribution.
Let us first illustrate this important fact with some temperature profiles and rubber friction results.

Fig. \ref{Slide_Temperature_profile_1.ps}
shows the measured temperature profile before run-in, after sliding $s= 3.5 \ {\rm m}$. The rubber block
is $L=2.5 \ {\rm cm}$ long in the sliding direction, $0.7 \ {\rm cm}$ high, and the nominal contact pressure
is $p=0.25 \ {\rm MPa}$.

Fig. \ref{Slide_Temperature_profile_2.ps}
shows the measured temperature profile after run-in. Here the run-in procedure consists of 10 repetitions, each involving sliding 3.5 m, i.e., a total of 35 m sliding distance.
Between each run there is a 10 second long waiting time. The temperature profile in Fig. \ref{Slide_Temperature_profile_2.ps} is obtained after the last repetition.

Note that in Fig. \ref{Slide_Temperature_profile_1.ps} the highest temperature occurs in a very localized region close to the leading edge.
Clearly before run-in the contact pressure is highly non-uniform with a maximum close to the leading edge.
During run-in the high temperature and stress at the leading edge result in local wear and finally the contact pressure at the sliding
interface becomes much more uniform, resulting in a more uniform temperature profile as seen in Fig. \ref{Slide_Temperature_profile_2.ps}. 
 
Fig. \ref{Slide_Laser_Profile_Rectangular-1a_and_2a.ps}
shows the measured profile of the rubber block (a) before run-in and
(b) after run-in (10 repetitions, each involving sliding 3.5 m). The run-in results in a rounding off of the leading edge of the
rubber block. The modification of the shape of rubber blocks during run-in has also been studied in detail
by Kr\"oger et al\cite{Kroger}. 

During run-in the rubber-road friction coefficient changes. 
Fig. \ref{page4_all.ps} shows the measured friction coefficients as a function of the number
of repetitions during run-in. Each repetition involves sliding
3.5 m at the sliding speeds (a) $v= 0.1$, (b) 0.5, and (c) $1 \ {\rm m/s}$.
Results are shown for the nominal contact pressures $0.05$ (blue lines),
$0.15$ (red), and $0.3 \ {\rm MPa}$ (green). Note that the friction coefficient tends to increase with increasing number of
run-in runs. During run-in the nominal contact pressure becomes more uniform and the local temperature increase in the region
where the friction force is generated is reduced, resulting in an increase in the friction force (see below). 

Let us now compare the measured temperature increase with the theory predictions. Unfortunately,
the topography of the asphalt road surface track has so far only been obtained using an optical method with limited
resolution, and the surface roughness power spectrum which enters in the theory calculations is therefore somewhat
uncertain, in particular for large wavenumber. The theory results which we now present are obtained from the equations
given in Ref. \cite{Fort}, and the thermal properties of the road and rubber are assumed the same as used in Ref. \cite{Fort}. 

Fig. \ref{1v.2T.p=0.05MPa.p=0.15.ps}
shows the measured maximal temperature on the rubber side wall (red data points) and the calculated
average (over the $y$-direction) rubber temperature at $x=L/2$ (middle of the nominal contact region) at the 
rubber-road interface (green symbols and lines).
Results are shown when the initial temperature equals $T=20$ and $40^\circ {\rm C}$, for the contact
pressures (a) $p=0.05$ and (b) $0.15 \ {\rm MPa}$. The block is $L=2.5 \ {\rm cm}$ long in the sliding direction
and the temperature refers to a sliding distance $s=3.5 \ {\rm m}$ after run-in. Note that for the lower nominal
contact pressure the theory agrees almost perfectly with the measured temperature increase, while some difference occurs for the
higher contact pressure.

Fig. \ref{Slide_Temperature_profile_2.ps} shows that the temperature in the nominal contact region is rather uniform except
close to the edges of contact. As we now show this indicates that the nominal contact pressure is not quite constant,
but probably increases from the trailing edge towards the leading edge, as indeed expected\cite{pressure}. 
The reason is that if the nominal contact pressure
would be constant one would expect a non-uniform temperature profile, where the temperature increases when moving from the leading edge
towards the trailing edge. This is illustrated in Fig. \ref{1x.2T.v=0.1m.per.s.ps}.

Fig. \ref{1x.2T.v=0.1m.per.s.ps} shows the calculated temperature $T(x)$ as a function of the position $x$ along the sliding
direction with $x=0$ at the leading edge and $x=L=2.5 \ {\rm cm}$ at the trailing edge. 
The temperature profile is obtained after sliding $s=3.5 \ {\rm m}$ at the nominal contact pressure
$p=0.15 \ {\rm MPa}$ and with the initial temperature $T=20^\circ {\rm C}$. 
The red line is the temperature in the macroasperity contact regions and the green line the temperature averaged 
over the $y$-direction, orthogonal to the sliding direction. Note that the temperature at the leading edge is close to the
road temperature, and maximal at the trailing edge. This is easy to understand: at the leading edge the rubber makes contact
with a road surface, which is at the same temperature as the surrounding air (equal to the initial temperature everywhere, 
in this case $T=20^\circ {\rm C}$). Since the temperature is (nearly) continuous in the road-rubber
asperity contact regions, the rubber temperature at the leading edge must be close to the initial
road (and air and rubber) temperature. At the exit of the contact the road asperities in contact with the rubber have heated up, 
resulting in the maximal rubber surface temperature at the exit of the contact. Such a non-uniform temperature profile is not observed in the
experiments, indicating a non-uniform nominal contact pressure. Presumably, after an even longer run-in time period, the contact pressure becomes
more uniform, with the (experimental) temperature profile $T(x)$ more similar to what is predicted theoretically (see also \cite{pressure}).
The rather small deviation between theory and experiment in Fig. \ref{1v.2T.p=0.05MPa.p=0.15.ps} is not unexpected considering that the nominal contact pressure profiles, 
which most likely prevail in the experiment, differ compared to the theory.

\vskip 0.3cm
{\bf 8 Implications for tire dynamics}

\begin{figure}
\includegraphics[width=0.45\textwidth,angle=0]{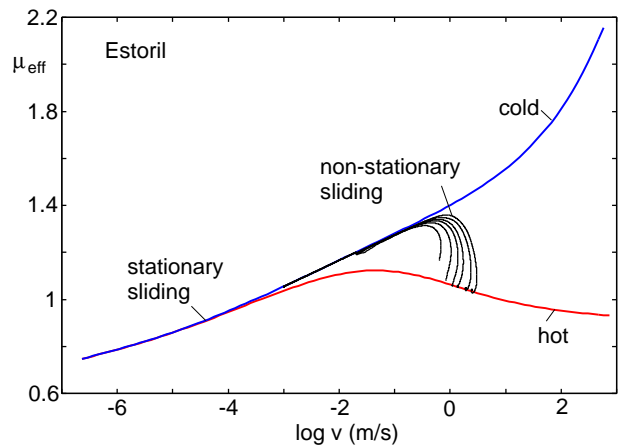}
\caption{\label{muslip}
Red and blue lines: the kinetic friction coefficient (stationary sliding) as a function of the logarithm 
(with 10 as basis) of the sliding velocity. The blue line denoted ``cold'' is without
the flash temperature while the red line denoted ``hot'' is with the flash temperature.
Black curves: the effective friction experienced by a tread block as it goes through the footprint.
For the car velocity $27 \ {\rm m/s}$ and for several 
slip values $0.005$, $0.0075$, $0.01$, $0.03$, $0.05$, $0.07$, and $0.09$. Note that the friction
experienced by the tread block first follows the ``cold'' rubber branch of the steady state kinetic friction coefficient and then, when the block has slip a distance of order the diameter of the macroasperity contact region,
it follows the ``hot'' rubber branch. 
}
\end{figure}

Rubber friction depends on the history of the sliding motion. This is mainly due to
the frictional heating: the temperature in the rubber-road asperity contact regions at time
$t$ depends on the sliding history for all earlier times $t' < t$. This memory effect is crucial for an
accurate description of rubber friction. We illustrate this effect in Fig. \ref{muslip}
for a rubber tread block sliding on an asphalt road surface\cite{Tire1}.
We show the (calculated) kinetic friction coefficient (including only the viscoelastic contribution)
for stationary sliding without (blue curve)
and including the flash temperature (red curve), as a function of the velocity $v$
of the bottom surface of the rubber block. The black curves show the effective friction during
non-stationary sliding experienced by a rubber tread block during braking at various slips
(slip values from 0.005 to 0.09). Note that because some finite sliding distance is necessary in order
to fully develop the flash temperature, the friction acting on the tread block initially
follows the blue curve corresponding to ``cold-rubber'' (i.e., negligible flash temperature).
Thus, it is not possible to accurately describe rubber friction
with just a static and a kinetic friction coefficient (as is often done even in advanced tire
dynamics computer simulation codes) or even with a function $\mu (v)$ which depends on the instantaneous sliding velocity
$v(t)$. Instead, the friction depends on $v(t')$ for all times $t' \le t$.

In tire applications, for slip of order $5-10\%$ and typical footprint length of order $10 \ {\rm cm}$,
the slip distance of a tread rubber block in the footprint will be of order $1 \ {\rm cm}$,
which typically is of order the diameters $D$
of the macro asperity contact regions. As discussed above, as long as the slip distance
$r(t)$ is small compared to $D$ one follows the cold rubber branch of the steady state
relation $\mu (v)$ so that $\mu(t) \approx \mu_{\rm cold}(v(t))$ for the slip distance
$r(t) << D$. When the tread block moves towards the end of the footprint the slip distance $r(t)$
may be of order (or larger than) $D$, and the friction will follow the hot branch of the
$\mu (v)$ relation, i.e., $\mu(t) \approx \mu_{\rm hot}(v(t))$ for $r(t) > D$.
We have found that the following (history dependent) friction law gives nearly the same result as the full
theory presented in Ref. \cite{Tire1,Tire2}:
$$\mu (t) = \mu_{\rm cold} (v,T) e^{-r(t)/r_0} +\mu_{\rm hot} (v,T) 
\left [1-e^{-r(t)/r_0}\right ]\eqno(2)$$
where $v=v(t)$ is the instantaneous sliding velocity, $r(t)$
the sliding distance at time $t$, and $r_0 \approx 0.2 D$. 
The background temperature $T$ in the surface region of the rubber 
is assumed to be independent of the spatial coordinate.

We will refer to (2) as the {\it cold-hot friction law}.
The length $D$ depends on the rubber compound and the road surface but is typically in the range
$D\approx 0.1-1 \ {\rm cm}$. Using the full friction theory one can easily calculate the functions
$\mu_{\rm cold} (v,T_0)$, $\mu_{\rm hot} (v,T_0)$, and the length $D$. 
So far, Eq. (2) has been tested when only the viscoelastic contribution is included, and the
equation may need modifications when also the adhesive contribution to the friction is 
included in the study.

\begin{figure}
\includegraphics[width=0.45\textwidth,angle=0]{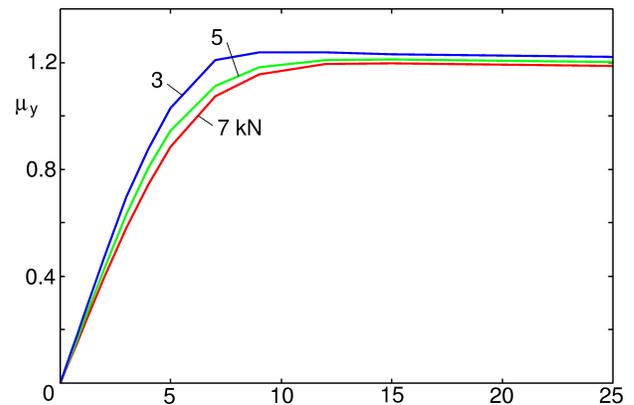}
\caption{\label{mu.slipangle.3000N.5000N.7000N.p0.3MPa.ellipse}
The $\mu$-slip angle curves for the elliptic footprint with the tire loads $F_{\rm N} = 3000$, $5000$ and $7000 \ {\rm N}$,
and the footprint pressure $p=0.3 \ {\rm MPa}$. 
For the rubber background temperature $T_0=80 \ ^\circ {\rm C}$
and the car velocity $27 \ {\rm m/s}$. 
}
\end{figure}

Let us illustrate the use if the cold-hot friction law (2), and the
influence of the flash temperature on tire dynamics, by presenting $\mu$-slip angle calculations for the 
simple 2-dimensional tire model described in Ref. \cite{Tire1,Tire2}.
Fig. \ref{mu.slipangle.3000N.5000N.7000N.p0.3MPa.ellipse} shows the 
$\mu$-slip angle curves for an elliptic footprint for the tire load $F_{\rm N} = 3000$, $5000$ and $7000 \ {\rm N}$,
and the footprint pressure $p=0.3 \ {\rm MPa}$. 
Note that as the load increases, the footprint becomes longer which results in a decrease in the maximum friction
coefficient, which agrees with experimental observations. 
This load-dependence is not due to an intrinsic pressure
dependence of the rubber friction coefficient (which was kept constant in our calculation), 
but a kinetic effect related to the build up of the flash
temperature in rubber road asperity contact regions during slip. To understand this in more detail, consider
again Fig. \ref{muslip}. 

The red and blue lines in Fig. \ref{muslip}
show the kinetic friction coefficient (stationary sliding) as a function of the logarithm
of the sliding velocity. The upper line denoted ``cold'' is without
the flash temperature while the lower line denoted ``hot'' is with the flash temperature.
The black curves show the effective friction experienced by a tread block as it goes through the footprint.
Note that the friction experienced by the tread block first follows the ``cold'' rubber branch, and then, when the block has slipped a 
distance of order the diameter $D$ of the macroasperity contact region,
it follows the ``hot'' rubber branch. Based on this figure it is easy to understand why the maximum
friction coefficient increases when the length of the footprint decreases: If $v_{\rm slip}$ is the
(average) slip velocity of the tread block, then in order to fully build up the flash temperature
the following condition must be satisfied:
$v_{\rm slip} t_{\rm slip} \approx D$, where $D$ is the diameter of the macroasperity
contact region. Since the time the rubber block stays in the
footprint $t_{\rm slip} = L/v_{\rm R}$ (where $L$ is the length of the footprint and $v_{\rm R}$
the rolling velocity) we get $v_{\rm slip} \approx v_{\rm R} (D/L)$. Thus, when the length
$L$ of the footprint decreases, the (average) slip velocity of the tread block in the footprint
can increase without the slip distance exceeding the diameter $D$ of the macro asperity contact
region. As a consequence, as $L$ decreases,
the tread block will follow the ``cold'' rubber branch of the (steady state) $\mu$-slip
curve to higher slip velocities before the flash temperature is fully developed, 
resulting in a higher (maximal) tire-road friction for a short 
footprint as compared to a longer footprint.

\vskip 0.3cm
{\bf 9 Summary and conclusion}

We have discussed several different origins for why $\mu$ could depend on the nominal pressure, thus violating the first Amonton's friction law. Possible explanations are:
(a) saturation of the contact area (and friction force) due to high nominal squeezing pressure, 
(b) non-linear viscoelasticity,
(c) non-randomness in the surface topography, and in particular the influence of the skewness of the surface roughness profile,
(d) adhesion and 
(e) frictional heating.

We have shown that for most cases the non-linearity in the $\mu(p)$ relation is mainly due to process (e) (frictional heating). This is for example the case for a tire in contact with the road surface where the contact area is usually just a small fraction of the nominal contact area. Here the energy dissipation in the contact regions between solids in sliding contact can result in high local temperatures which may strongly affect the area of real contact and the friction force (and the wear rate). This is usually observed for rubber sliding on road surfaces at speeds above $1 \ {\rm mm/s}$. In this paper we have presented and compared numerical results to experimental data for the temperature increase of a rubber block in sliding contact with a randomly rough surface. We observe good agreement between the calculated and measured temperature increase.

Rubber friction depends very sensitively on the rubber temperature shifting the viscoelastic and the contact area contribution towards higher sliding speeds with increasing temperature, 
usually resulting in a decrease of the friction coefficient. When the load (or nominal contact pressure) increases, 
the effect of frictional heating increases as well, resulting in even higher temperatures and a non-linear dependency of the friction force on the applied normal load.

However, depending on the system studied, the non-linear relation of $\mu(p)$ can also be attributed to a saturation of contact area with normal load or to skewness of the roughness profile affecting the rubber friction. The area of real contact between elastic (or viscoelastic) solids with nominally flat, but randomly rough surfaces is usually proportional to the applied normal force as long as the area of real contact is less that $\sim 30\%$ of the nominal contact area. When the load (or nominal force) becomes so high that the contact area exceeds $\sim 30\%$ of the nominal contact area, the friction force starts to depend non-linearly on the load.

An increase of the normal load on surfaces with a skewed height profile will make the rubber surface penetrates deeper into the roughness profile, where it will experience roughness components with different statistical properties.
This will influence the area of contact and also the viscoelastic contribution to rubber friction. The effect gets enhanced when frictional heating becomes high enough, 
leading to a softening of the rubber, an even deeper penetration of the rubber into the roughness profile, and (for surfaces with negative skewness) a reduction of the viscoelastic contribution to the friction.

\end{document}